\documentclass[10pt,twocolumn,letterpaper]{article}

\usepackage{cvpr}              

\usepackage{makecell}
\usepackage[flushleft]{threeparttable}
\usepackage{arydshln}
\usepackage{amssymb}

\usepackage[dvipsnames]{xcolor}

\newcommand{\zhihao}[1]{#1}

\def\encoder{ f_\text{enc} }
\def\decoder{ f_\text{dec} }

\definecolor{cvprblue}{rgb}{0.21,0.49,0.74}
\usepackage[pagebackref,breaklinks,colorlinks,citecolor=cvprblue]{hyperref}

\def\paperID{1616} 
\def\confName{CVPR}
\def\confYear{2024}

\title{Towards Backward-Compatible Continual Learning of Image Compression}

\author{
Zhihao Duan\textsuperscript{1} \quad
Ming Lu\textsuperscript{2} \quad
Justin Yang\textsuperscript{1} \quad
Jiangpeng He\textsuperscript{1}\(^{\dagger}\) \quad
Zhan Ma\textsuperscript{2} \quad
Fengqing Zhu\textsuperscript{1}
\\
\textsuperscript{1} Purdue University, West Lafayette, Indiana, U.S.A. \\
\textsuperscript{2} Nanjing University, Nanjing, Jiangsu, China \\
{\tt\small \{duan90, yang1834, he416, zhu0\}@purdue.edu, \{minglu, mazhan\}@nju.edu.cn}
}

\begin{document}

\maketitle
\renewcommand{\thefootnote}{\fnsymbol{footnote}} 
\footnotetext[2]{Corresponding author \& Project lead}

\begin{abstract}
\zhihao{
This paper explores the possibility of extending the capability of pre-trained neural image compressors (\eg, adapting to new data or target bitrates) without breaking backward compatibility, the ability to decode bitstreams encoded by the original model.
We refer to this problem as continual learning of image compression.
Our initial findings show that baseline solutions, such as end-to-end fine-tuning, do not preserve the desired backward compatibility.
To tackle this, we propose a knowledge replay training strategy that effectively addresses this issue.
We also design a new model architecture that enables more effective continual learning than existing baselines.
Experiments are conducted for two scenarios: data-incremental learning and rate-incremental learning.
The main conclusion of this paper is that neural image compressors can be fine-tuned to achieve better performance (compared to their pre-trained version) on new data and rates without compromising backward compatibility.
Our code is available at \href{https://gitlab.com/viper-purdue/continual-compression}{this link}.
}
\end{abstract}

\section{Introduction}
\label{sec:cloc_intro}


Recent years have witnessed the rapid development of deep learning-based image compression.
Most existing research in this field considers the \textit{offline learning} setting, \ie, once a neural network model is trained, its parameters are fixed and kept unchanged when deployed.
However, real-world applications are often complex and dynamic, and an ideal compressor should be capable of being incrementally learned and adapted to various scenarios.
For example, consider an image storage application with a compressor pre-trained on natural images for certain predefined target bitrates.
As new image sources (\eg, microscopy, remote sensing, and human face images) are encountered, one may want to update the compressor to improve its performance on the new data and to support different target rates.
This raises an interesting question: \textit{can neural network-based image compressors be learned continually, and if so, would it bring performance benefits compared to the pre-trained model?}

\begin{figure}[t]
    \centering
    \begin{subfigure}{\linewidth}
        \centering
        \includegraphics[width=0.8\linewidth]{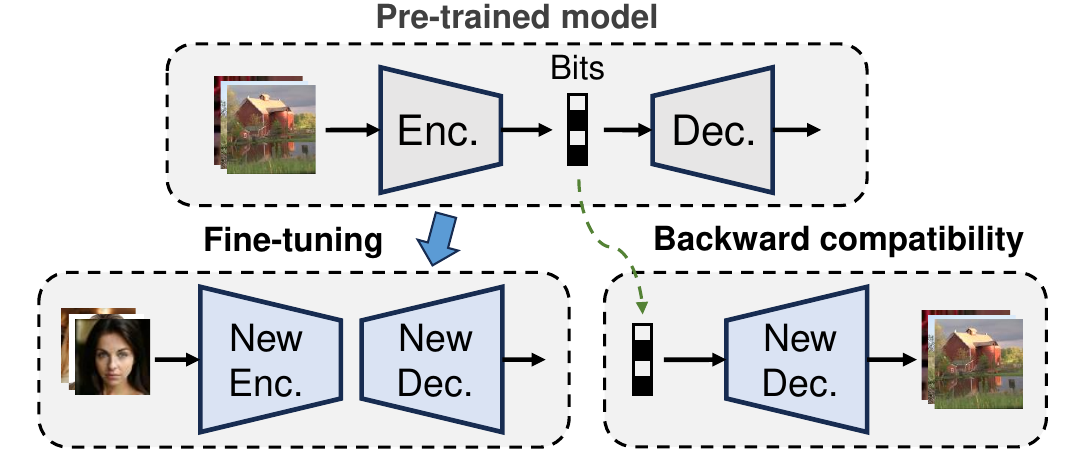}
        \caption{Continual learning of image compression}
        \vspace{0.12cm}
        \label{fig:cloc_intro_problem}
    \end{subfigure}
    \begin{subfigure}{\linewidth}
        \centering
        \includegraphics[width=0.92\linewidth]{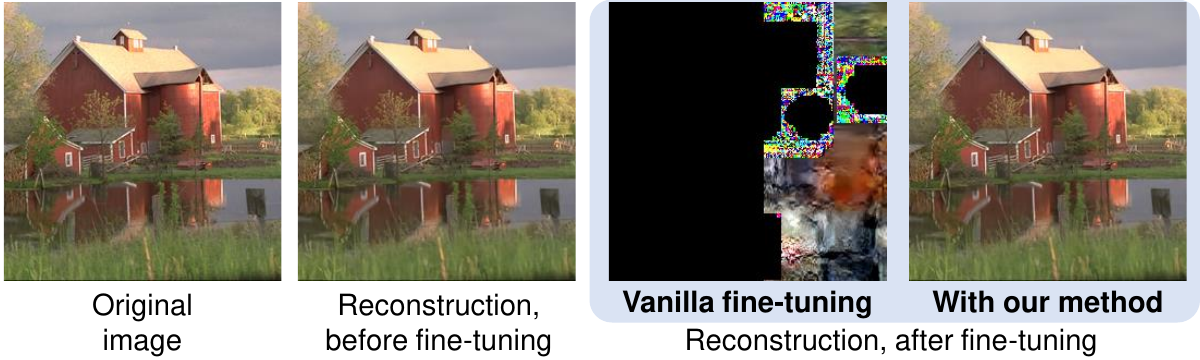}
        \vspace{-0.04cm}
        \caption{The backward compatibility problem.
            Once a neural compressor (in this experiment, \cite{minnen2018joint}) is fine-tuned, it can no longer decode the bitstream produced by its original version. Our method addresses this issue.
        }
        \label{fig:cloc_intro_finetune}
    \end{subfigure}
    \hfill
    \vspace{-0.6cm}
    \caption{
        \zhihao{The goal of this work is to fine-tune pre-trained learned image compressors with new data or new rates while preserving backward compatibility (\cref{fig:cloc_intro_problem}). Baseline methods are backward incompatible, while ours is effective (\cref{fig:cloc_intro_finetune}).}
    }
    \label{fig:cloc_intro}
    \vspace{-0.32cm}
\end{figure}

One might assume that simply fine-tuning a pre-trained model would be sufficient.
Yet, doing so disrupts the model's \textit{backward compatibility} (\cref{fig:cloc_intro_finetune}), \ie, the ability to decode bitstreams produced by the original model, due to a mismatch between the encoder (pre-trained model) and decoder (fine-tuned model).
Maintaining backward compatibility is crucial, as failing to do so renders existing bitstreams in the storage (or sent from other devices) inaccessible.
It is worth noting that this backward compatibility problem is different from the well-known problem of catastrophic forgetting in neural networks~\cite{kirkpatrick2017catastrophic_forgetting}.
The unique properties of compression, including the sender-receiver relationship and the existence of entropy coding, set it apart from other image processing/vision tasks.
Therefore, existing continual learning methods for vision tasks~\cite{lange2022tpami_cl_survey, wang2023comprehensive_cl_survey} cannot be applied as is, and new strategies must be developed for compression to maintain the decoder's backward compatibility when adapting to new data and rates.

To achieve backward compatibility, the most straightforward way is to modify only the encoding process of a pre-trained model when adapting to new data or new rates~\cite{campos2019content_adaptive, yang2020improving, gao2022code_editing}.
This strategy resembles the common practice in traditional codecs, where there is often a standardized decoder, but various encoders can be designed to accommodate different applications.
Despite its simplicity, keeping the decoder unchanged hinders the model's ability to adapt to new data and rates, leading to suboptimal performance.

This work shows that it is possible to continually train both the encoder and decoder of neural compressors while maintaining backward compatibility.
We begin by noticing that as long as the entropy model of a compressor is kept unchanged, it can decode the old bitstreams and obtain the latent representations.
Based on this observation, we propose a knowledge replay training scheme that can be used to train the encoder and decoder networks without breaking backward compatibility.
We also design a model architecture where the entropy model consumes only a small amount of parameters, allowing most model parameters to be learnable in fine-tuning.
\zhihao{
We formulate two experimental scenarios: data-incremental learning and rate-incremental learning.
Experimental results demonstrate that our proposed methods enable neural image compressors to obtain improved performance on new data and new rates without breaking backward compatibility.
}


To summarize, our contributions are as follows:
\begin{itemize}
    \item We propose a knowledge replay-based training strategy that can be used to train neural image compressors incrementally without breaking backward compatibility;
    \item We design a neural network architecture targeting effective continual learning of image compression;
    \item We formulate two continual learning scenarios for image compression: data-incremental learning and rate-incremental learning. Experimental results show that our method outperforms baseline approaches in both cases.
\end{itemize}

\section{Background and Related Work}
\label{sec:cloc_related}

\subsection{Learned lossy image compression (LIC)}
Most learning-based methods for lossy image compression can be interpreted using the \textit{entropy-constrained non-linear transform coding} framework~\cite{balle2021nonlinear}.
Let $X \sim p_\text{data}$ denote data samples with an underlying data distribution.
In this framework, a neural network encoder $\encoder$ maps $X$ to a latent variable $Z \triangleq \encoder(X)$, and a neural network decoder $\decoder$ maps $Z$ back to a reconstruction $\hat{X} \triangleq \decoder(Z)$.
A learned probability distribution $p_Z$, also known as the \textit{entropy model}, is used to model the marginal distribution of $Z$.
The learning objective is to minimize the rate-distortion (R-D) loss:
\begin{align}
    \label{eq:cloc_background_rd_loss}
    \min \, \mathbb{E}_{X \sim p_\text{data}} \left[ - \log_2 p_Z(Z) + \lambda \cdot d(X, \hat{X}) \right],
\end{align}
where $d$ is a distortion metric (\eg, mean squared error), $\lambda$ is the Lagrange multiplier trading off between rate and distortion, and the minimization is over the network parameters of $\encoder$, $\decoder$, and $p_Z$.
This framework has also been extended to \textit{variable-rate compression}~\cite{choi2019var_rate_conditional_ae, duan2023qarv}, where the encoder, decoder, and entropy model are conditioned on $\lambda$.
During variable-rate training, the model parameters are optimized for various $\lambda$ sampled from a distribution $p_\Lambda$:
\begin{align}
    \label{eq:cloc_background_rd_loss_var_rate}
    \min & \, \mathbb{E}_{X \sim p_\text{data}, \Lambda \sim p_\Lambda} \left[ - \log_2 p_Z(Z | \Lambda) + \Lambda \cdot d(X, \hat{X}) \right]
    \\
    & \text{where } Z \triangleq \encoder(X; \Lambda), \quad \hat{X} \triangleq \decoder(Z; \Lambda).
\end{align}

Existing research in LIC can be categorized into several groups.
A major group focuses on designing expressive architectures for $\encoder$ and $\decoder$, such as convolutional and transformer-based ones~\cite{cheng2020cvpr, chen2021nlaic, xie2021invertible, qian2021global_ref, gao2021back_projection, lu2022tic, ma2022tpami_iwave, zou2022stf, he2022elic, lu2022tinylic, duan2023qarv, liu2023cvpr_tcm, jiang2023mlic, yang2023iccv_shallow_decoder}.
Another line of research lies in designing the entropy model $p_Z$, such as autoregressive~\cite{minnen2018joint, minnen2020channelwise, he2021checkerboard, qian2022entroformer} and hierarchical models~\cite{balle18hyperprior, hu2022tpami_benchmark, duan2023qres, duan2023qarv}.
Other research includes, \eg, quantization methods~\cite{yang2020bayes_quantization, guo2021soft_then_hard, alaaeldin2023product_quantized, feng2023nvtc, zhang2023lvqac} and variable-rate compression methods~\cite{choi2019var_rate_conditional_ae, chen2020var_rate_quality_scaling, yang2021slimmable, song2021qmap, cai2022var_rate_invertible_activation, lee2022nips_selective_compression}.
To our knowledge, none of these existing methods are designed to work with continual learning as in our work.

The most related line of research to this paper is \textit{\textbf{content-adaptive image compression}}, where the goal is to adapt a compressor to new images or new target rates in a per-image fashion.
Solutions to this problem include encoder-side optimization and decoder-side adaptation.
Encoder-side optimization methods~\cite{campos2019content_adaptive, yang2020improving, gao2022code_editing} directly optimize the R-D objective w.r.t. $Z$ during encoding.
Decoder-side adaptation methods~\cite{pan2022content_adaptive, tsubota2023wacv_content_adaptive, shen2023dec_adapter}, on the other hand, include parameter-efficient neural network modules in the bitstream, which are executed on the decoder side to improve decoding.
Among them, many methods require computationally expensive, iterative optimization during encoding.

\zhihao{
The scope of this paper is distinct from content-adaptive image compression in several ways:
(a) our goal is to incrementally train the compressor parameters in-place without introducing additional parameters, and (b) as opposed to per-image optimization during testing, our method employs a one-time training procedure and induces no additional computational cost at test time.
Our research is complementary to content-adaptive image compression, and they can be combined to further improve the performance.
}

\subsection{Knowledge replay in continual learning}
Continual learning has been widely studied for image classification~\cite{lange2022tpami_cl_survey} and semantic segmentation~\cite{cermelli2020modeling_seg}, which aim to learn a sequence of new tasks incrementally without forgetting the previously learned knowledge.
Among existing methods for continual learning, \textit{replay-based} methods~\cite{rebuffi2017icarl, liu2020mnemonics, lopez2017gradient_gem} have emerged as particularly effective, which applies an additional memory buffer to store exemplar data from learned tasks and then integrate with new task data to perform knowledge rehearsal during continual learning.
Our proposed knowledge replay method adopts a similar principle.
However, image compression contains unique challenges distinct from typical computer vision tasks, making existing continual learning strategies inapplicable to deploy in image compression.
Thus, a tailored knowledge replay strategy is required for our problem scenario.





\section{Problem Description}
\label{sec:cloc_problem}

In this section, we formulate the problem (\cref{sec:cloc_problem_statement}), discuss the backward compatibility issue (\cref{sec:cloc_problem_backward_compatibility}), and motivate the design of our method (\cref{sec:cloc_problem_em_size}).

\begin{figure}[t]
    \centering
    \begin{subfigure}{0.48\linewidth}
        \includegraphics[width=\linewidth]{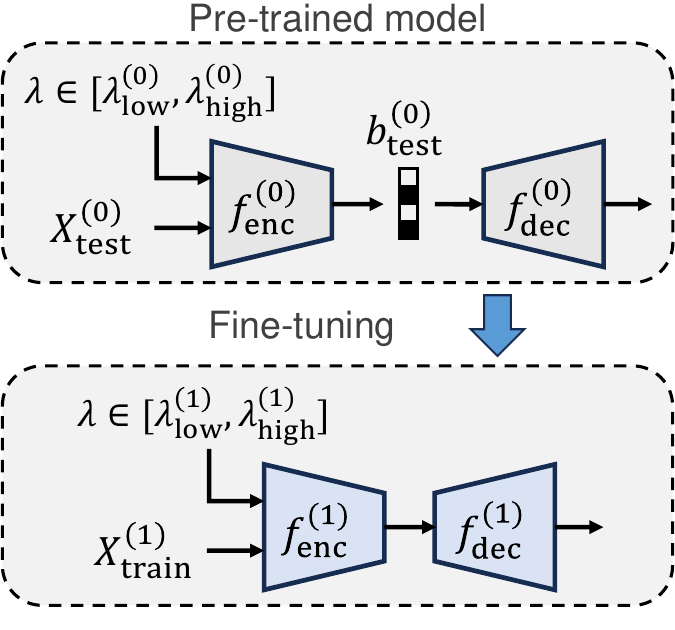}
        \caption{Continual learning}
        \label{fig:cloc_problem_statement_stages}
    \end{subfigure}
    \hfill
    \begin{subfigure}{0.48\linewidth}
        \includegraphics[width=\linewidth]{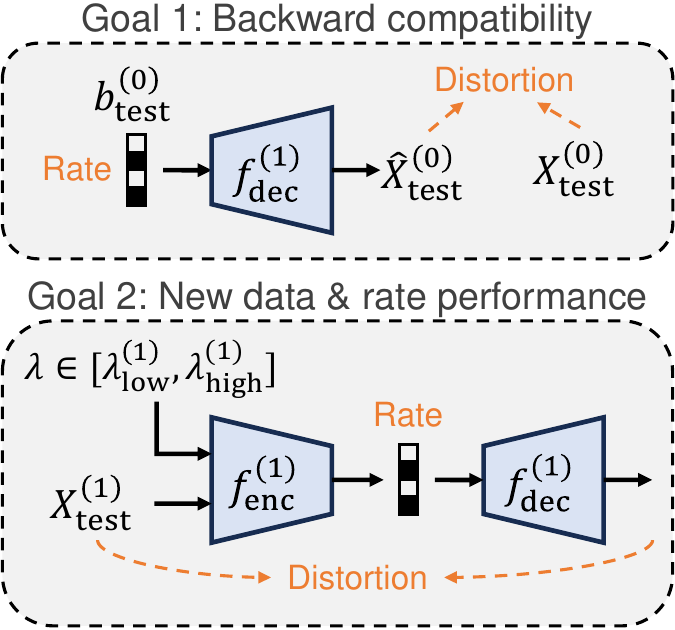}
        \caption{Evaluation}
        \label{fig:cloc_problem_statement_eval}
    \end{subfigure}
    \vspace{-0.2cm}
    \caption{
        Problem definition. \cref{fig:cloc_problem_statement_stages} shows the fine-tuning process of a pre-trained model, and \cref{fig:cloc_problem_statement_eval} shows the two evaluation criteria: backward compatibility and new data/rate performance. Entropy models are kept frozen and omitted in the figures.
    }
    \label{fig:cloc_problem_statement}
\end{figure}

\subsection{Continual learning for compression}
\label{sec:cloc_problem_statement}

Assume that we have a pre-trained variable-rate model with supported $\lambda \in [\lambda^\text{(0)}_\text{low}, \lambda^\text{(0)}_\text{high}]$.
We use $\encoder^\text{(0)}, \decoder^\text{(0)}, p_Z^\text{(0)}$ to denote the pre-trained model's encoder, decoder, and entropy model, respectively.
Also, assume that we have used the pre-trained model to compress a test set of images $X^\text{(0)}_\text{test}$ and obtained the corresponding bitstreams $b^\text{(0)}_\text{test}$, as illustrated in \cref{fig:cloc_problem_statement_stages} (top).
This situation well-silumates a typical image storage application with a learned image compressor.

We now aim at fine-tuning the model with new data $X^\text{(1)}_\text{train}$ and a new rate range determined by $[\lambda^\text{(1)}_\text{low}, \lambda^\text{(1)}_\text{high}]$.
Note that the new data and rate range may or may not be the same as the old ones.
Similarly, let $\encoder^\text{(1)}, \decoder^\text{(1)}, p_Z^\text{(1)}$ denote the new model components, as illustrated in \cref{fig:cloc_problem_statement_stages} (bottom).
We expect the new model to achieve two goals:
\begin{enumerate}
    \item \textbf{Backward compatibility:} The new model should be capable of decoding the bitstreams produced by the old model (\cref{fig:cloc_problem_statement_eval}, top).
    \item \textbf{New-data (or new-rate) performance}: The new model should perform better than the old one on new data and new rates (\cref{fig:cloc_problem_statement_eval}, bottom).
\end{enumerate}

\subsection{Backward compatibility of entropy decoding}
\label{sec:cloc_problem_backward_compatibility}

As mentioned in \cref{fig:cloc_intro}, fine-tuning the model end-to-end breaks the backward compatibility of the decoder.
The primary reason lies in entropy coding: range-based entropy coding algorithms~\cite{rissanen1979arithmetic_coding, duda2015pcs_ans}, which are commonly used in modern neural compressors, are known to be sensitive to the probability distribution of the encoded symbols.
As the new entropy model $p_Z^\text{(1)}$ is different from the old one $p_Z^\text{(0)}$, the new model is not able to perform entropy coding correctly using the old bitstreams $b^\text{(0)}_\text{test}$, and thus the correct (quantized) latent variables $Z^\text{(0)}_\text{test}$ cannot be obtained.
In fact, recent research~\cite{balle2018integer, koyuncu2022lic_quantization, tian2023effortless} has shown that even a small change (\eg, a floating point round-off error) in the entropy model can lead to failure in entropy decoding.

\begin{figure}[t]
    \centering
    \includegraphics[width=0.64\linewidth]{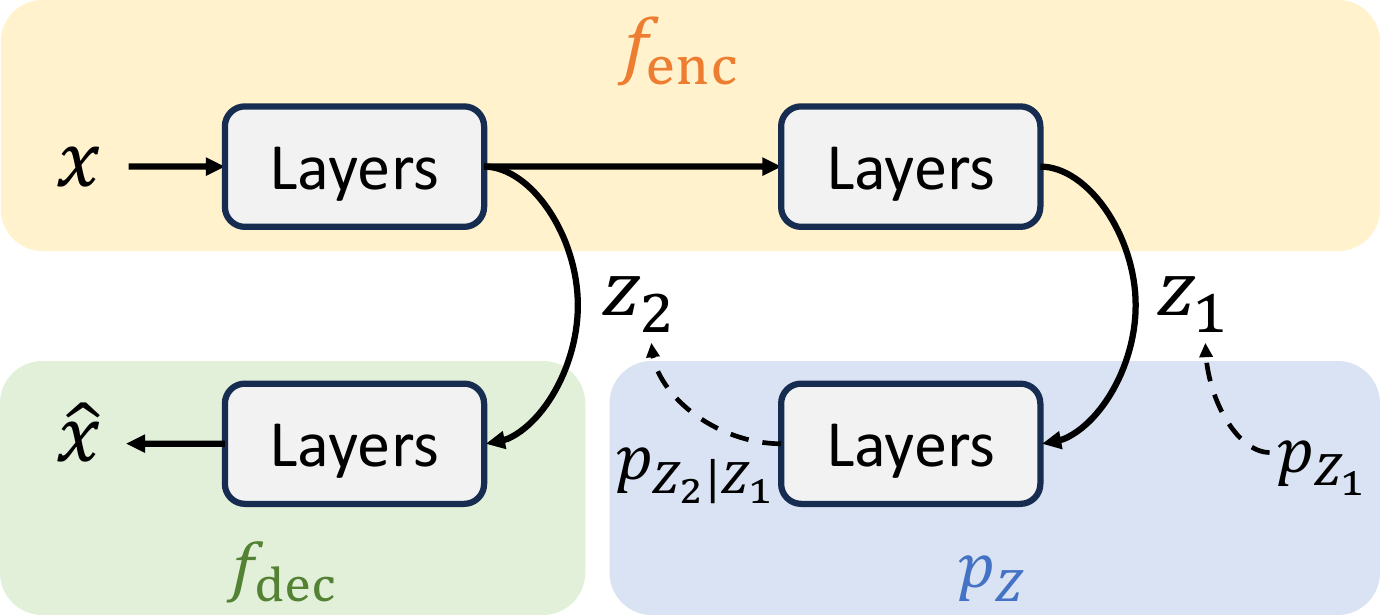}
    \vspace{-0.2cm}
    \caption{
        The Hyperprior model architecture~\cite{balle18hyperprior}, where the layers are categorized into three groups: $\encoder$, $\decoder$, and $p_Z$.
    }
    \label{fig:cloc_problem_hyperprior}
\end{figure}
\begin{table}[t]
    \centering
    \footnotesize
    \begin{tabular}{l|c:ccc}
        \hline
                                        & \multicolumn{4}{c}{\textbf{Number of parameters}}  \\ \hline
                                        & Total & $\encoder$ & $\decoder$  & $p_Z$  \\ \hline
        M-S Hyp.~\cite{minnen2018joint} & 17.6M & 5.9M  & 2.5M      & 8.2M (47\%)   \\
        GMA~\cite{cheng2020cvpr}        & 26.6M & 5.8M  & 11.7M     & 9.0M (34\%)   \\
        ELIC~\cite{he2022elic}          & 33.8M & 9.7M  & 7.3M      & 16.7M (49\%)  \\
        STF~\cite{zou2022stf}           & 99.9M & 11.3M & 7.1M      & 81.4M (81\%)  \\
        TCM~\cite{liu2023cvpr_tcm}      & 45.2M & 3.3M  & 6.8M      & 35.2M (78\%)  \\
        MLIC++~\cite{jiang2023mlic_pp}  & 116.7M& 6.3M  & 12.0M     & 98.4M (84\%)  \\ \hline
        Our model                       & 35.5M & 18.7M & 11.9M     & \textbf{4.9M (14\%)}   \\ \hline
    \end{tabular}
    \vspace{-0.1cm}
    \caption{Many existing methods employ a parameter-expensive entropy model. We propose an architecture with a lightweight entropy model, which makes continual learning more effective (since more parameters are learnable in the fine-tuning phase).}
    \label{table:cloc_num_params}
\end{table}

\subsection{Freezing the entropy model during fine-tuning}
\label{sec:cloc_problem_em_size}

To avoid the aforementioned issue, the entropy model $p_Z$ needs to be kept unchanged throughout fine-tuning.
Then, the latent variables $Z^\text{(0)}_\text{test}$ are guaranteed to be recovered losslessly from the old bitstreams, and the problem reduces to continually learning the decoder $\decoder$ without forgetting the old knowledge (\ie, decoding $Z^\text{(0)}_\text{test}$).
With our proposed training strategy (\cref{sec:cloc_method_knowledge_replay}), we show that this can be achieved for many existing neural compressors.

We also notice that the entropy model $p_Z$ is often the largest component in many existing compressors, most of which are based on the Hyperprior model~\cite{balle18hyperprior,minnen2018joint} (\cref{fig:cloc_problem_hyperprior}).
We found that their entropy model takes up a significant portion (\eg, 84\% for MLIC++\cite{jiang2023mlic_pp}) of the model parameters, as shown in \cref{table:cloc_num_params}.
Consequently, a large proportion of model parameters need to be fixed during continual learning, which may impair the new-data (or new-rate) performance.
Motivated by this, we design a model architecture (\cref{sec:cloc_method_architecture}) that employs a lightweight entropy model (\cref{table:cloc_num_params}, last row), which makes continual learning more effective.



\section{Method}
\label{sec:cloc_method}

\zhihao{
As just mentioned, our methods include two independent components: the knowledge replay-based training strategy (\cref{sec:cloc_method_knowledge_replay}) and a neural network architecture specifically designed for continual learning (\cref{sec:cloc_method_architecture}).
We now present them sequentially in detail.
}


\subsection{Continual learning with knowledge replay}
\label{sec:cloc_method_knowledge_replay}


Following the notation introduced in \cref{sec:cloc_problem_statement}, we denote a pre-trained model as $\encoder^{(0)}, \decoder^{(0)}, p_Z$, which are trained on data $X_\text{train}^{(0)}$ with $\lambda \in [\lambda^{(0)}_\text{low}, \lambda^{(0)}_\text{high}]$.
We drop the superscript for $p_Z$ since it is kept frozen throughout fine-tuning.
Inspired by the idea of knowledge rehearsal with exemplars in class-incremental learning methods~\cite{rebuffi2017icarl}, we use the old training data $X_\text{train}^{(0)}$ and the old encoder $\encoder^{(0)}$ to perform ``\textit{knowledge replay}'' in the fine-tuning process.
To this end, we store $\encoder^{(0)}$ along with $X_\text{train}^{(0)}$ within a dedicated memory buffer before the fine-tuning stage.
Note that we do not pose restrictive assumptions on training resources and allow access to the entire training set $X_\text{train}^{(0)}$ during fine-tuning.


In the fine-tuning process, our knowledge replay-based training objective contains two terms.
The first term, $\ell_\text{new}$, is the standard R-D loss for the new training data $X_\text{train}^{(1)}$ with the new $\lambda$ value range $[\lambda^{(1)}_\text{low}, \lambda^{(1)}_\text{high}]$:
\begin{align}
    \label{eq:cloc_method_loss_new}
    \ell_\text{new} & \triangleq \mathbb{E} \left[  R^{(1)} + \Lambda^{(1)} \cdot D^{(1)} \right], \quad \text{where }
    \\
    R^{(1)} & \triangleq -\log_2 \, p_Z(Z^{(1)} | \Lambda^{(1)})
    \\
    D^{(1)} & \triangleq d(X_\text{train}^{(1)}, \hat{X}_\text{train}^{(1)}).
\end{align}
In \eqref{eq:cloc_method_loss_new}, the expectation is w.r.t. $X_\text{train}^{(1)}$ and $\Lambda^{(1)}$, where $\Lambda^{(1)}$ is a random variable with the support being $[\lambda^{(1)}_\text{low}, \lambda^{(1)}_\text{high}]$.
Its probability density, $p_\Lambda^{(1)}$, controls how $\lambda$ is sampled during training.
Intuitively, minimizing $\ell_\text{new}$ adapts the model parameters to new data and new rates, but it is not sufficient to maintain backward compatibility.

The other term in our loss function encourages backward compatibility of the model parameters through knowledge replay of the old data $X_\text{train}^{(0)}$ and the old encoder $\encoder^{(0)}$.
Specifically, we use $\encoder^{(0)}$ to encode $X_\text{train}^{(0)}$, which gives the corresponding (quantized) latent variables $Z_\text{train}^{(0)}$.
Then, the current decoder $\decoder^{(1)}$ decodes $Z_\text{train}^{(0)}$, and the reconstruction $\hat{X}_\text{train}^{(0)}$ is compared with $X_\text{train}^{(0)}$ to compute the knowledge replay loss function, $\ell_\text{KR}$, defined as:
\begin{align}
    \ell_\text{KR} & \triangleq \mathbb{E} \left[ \Lambda^{(0)} \cdot D^{(0)} \right], \quad \text{where }
    \\
    D^{(0)} & \triangleq d \left( X_\text{train}^{(0)}, \, \decoder^{(1)}\left(\encoder^{(0)}(X_\text{train}^{(0)})\right) \right).
\end{align}
The expectation is w.r.t. $X_\text{train}^{(0)}$ and $\Lambda^{(0)}$, where $\Lambda^{(0)} \sim p_\Lambda^{(0)}$ controls how $\lambda$ is sampled during knowledge replay.
In our experiments, we choose $p_\Lambda^{(0)}$ to be the same as the one used in pre-training.
Note that there is no rate term in $\ell_\text{KR}$ since the replayed encoder $\encoder^{(0)}$ is fixed, and thus the rate term is constant w.r.t. the model parameters being trained.
By replaying the old data and the old encoder, the decoder network effectively retains backward compatibility, as shown in our experiments (\cref{sec:cloc_exp_baseline_comparison}).

\begin{figure}[t]
    \centering
    \includegraphics[width=0.8\linewidth]{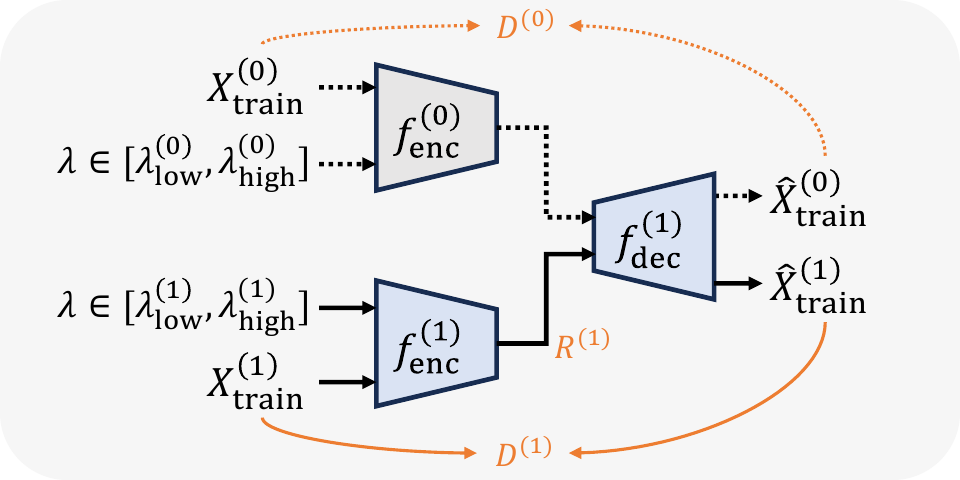}
    \vspace{-0.2cm}
    \caption{
        Our knowledge replay-based training strategy contains two components: a distortion loss that encourages backward compatibility (top, dashed lines), and the standard R-D loss for new data and new rates (bottom, solid lines). The entropy model $p_Z$ is kept frozen and is omitted in the figure.
    }
    \label{fig:cloc_method_kr}
\end{figure}

\cref{fig:cloc_method_kr} illustrates our knowledge replay strategy for one training iteration.
Our training objective is to minimize a weighted summation of the two terms, $\ell_\text{new}$ and $\ell_\text{KR}$, with a scalar hyperparameter $\alpha \in [0,1]$:
\begin{align}
    \min \, (1 - \alpha) \cdot \ell_\text{new} + \alpha \cdot \ell_\text{KR}.
    \label{eq:cloc_method_kr_loss}
\end{align}
The hyperparameter $\alpha$ controls the weight of replayed data in the training objective and can be tuned to achieve a desired trade-off between backward compatibility and new data/rate performance (shown in \cref{sec:cloc_exp_study_replay}).
Our knowledge replay strategy is general and can be applied to various model architectures, enabling backward-compatible continual learning of those models (shown in \cref{sec:cloc_exp_baseline_comparison}).

\subsection{Model architecture}
\label{sec:cloc_method_architecture}
\begin{figure}[t]
    \centering
    \includegraphics[width=0.8\linewidth]{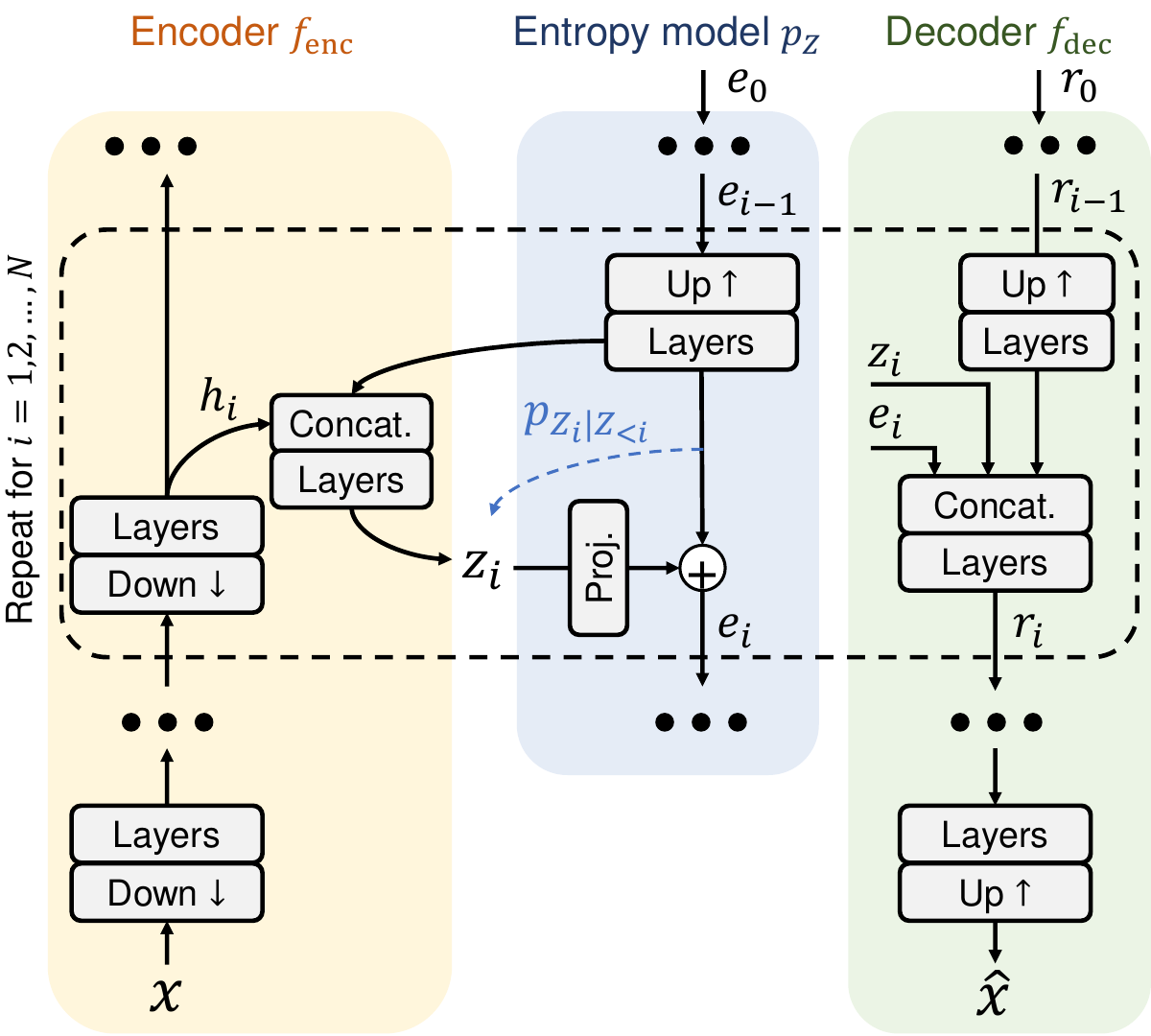}
    \caption{
        Our model adopts the hierarchical residual coding architecture~\cite{zhou2022mcquic, feng2023nvtc, duan2023qarv} but decouples the entropy model ($p_Z$) and the decoder ($\decoder$) into two branches.
        \textit{Up $\uparrow$} denotes upsampling, \textit{Down $\downarrow$} denotes down-sampling, and \textit{Proj.} denotes a linear projection layer.
        Detailed layer configurations are in Appendix, \cref{sec:cloc_app_model_arch}
    }
    \label{fig:cloc_method}
\end{figure}

Since freezing the entropy model is necessary for backward compatibility, we propose a model architecture that employs a lightweight entropy model by design so that most parameters in the model can stay learnable.
An overview of the architecture is shown in \cref{fig:cloc_method}.
Our model is inspired by the hierarchical residual coding architecture~\cite{zhou2022mcquic, feng2023nvtc, duan2023qarv}, but we decouple the entropy model ($p_Z$) and the decoder ($\decoder$) into two separate branches, resulting in a reduced entropy model size.
We now describe each component of the model in detail.

\textbf{Encoding:}
The encoding process involves a bottom-up pass through the encoder $\encoder$ and a top-down pass through the entropy model $p_Z$.
Given an input image $x$, encoding begins with $\encoder$, a neural network consisting of a sequence of downsampling and residual layers that produces a hierarchy of image features (denoted as $h_i$ in the figure).
Specifically, for an input image with $256 \times 256$ pixels, $\encoder$ produces $N = 4$ features with spatial dimentions $32 \times 32$, $16 \times 16$, $8 \times 8$, and $4 \times 4$.
All layers are convolutional, so the spatial dimensions scale accordingly for images of different sizes.
Then, the entropy model $p_Z$ starts with a constant $e_0$ and iteratively updates it using the features $h_i$ from $\encoder$.
In each stage, $e_{i-1}$ is upsampled to the same spatial dimension as $h_i$ and concatenated with $h_i$.
The concatenated features are then fed into a sequence of layers to produce $z_i$, the latent variable (which will be entropy coded) for the $i$-th stage.
$z_i$ is then aggregated with the upsampled $e_{i-1}$ through a linear layer and addition operation, and the result is denoted as $e_i$ and passed to the next stage.
Note that in each stage, the entropy model also estimates the probability distribution of $z_i$ given $z_{<i}$ (with $z_{<i} \triangleq \{ z_1, ..., z_{i-1} \}$), which is used for entropy coding.

\textbf{Probabilistic model, quantization, and entropy coding} are performed in the same way as for the discretized Gaussian model in Hyperprior-based methods~\cite{balle18hyperprior, minnen2018joint}.
As a brief recap, the entropy model predicts a mean $\mu_i$ and a scale $\sigma_i$ for each latent variable $z_i$, and the probability model for $z_i$ is a discretized Gaussian distribution:
\begin{align}
    p_i(z_i) &\triangleq p_{Z_i \mid Z_{<i}} (z_i \mid z_{<i})
    \\ &= \int_{z_i - 0.5}^{z_i + 0.5} \mathcal{N} (t ; \, \mu_i, \sigma_i^2) \, dt,
\end{align}
where the dependence on $z_{<i}$ is through $\mu_i$ and $\sigma_i$.
During testing, the residual between $z_i$ and $\mu_i$ is quantized to the nearest integer, and during training, quantization is simulated by additive uniform noise.
Each stage $i$ produces a separate bitstream corresponding to $z_i$ (using the asymmetric numeral systems~\cite{duda2015pcs_ans} algorithm), and all stages are executed sequentially for $i = 1,..., N$.

\begin{figure}[t]
    \centering
    \includegraphics[width=0.64\linewidth]{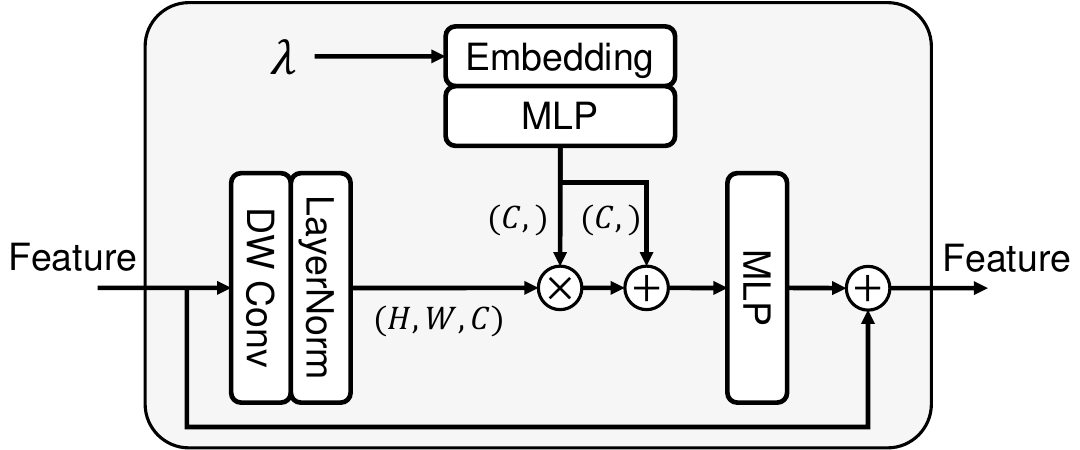}
    \caption{Each layer in our model is a ConvNeXt block~\cite{liu2022convnext} conditioned on $\lambda$ through an affine transformation. In the figure, $(H,W,C)$ denotes height, width, and channel dimensions.}
    \label{fig:cloc_method_layer}
\end{figure}

\textbf{Decoding:}
Given encoded bitstreams, the decoding process mirrors the encoding process in reverse.
Firstly, the entropy model is executed top-down to iteratively predict $p_{Z_i \mid Z_{<i}}$, based on which the bitstreams are entropy-decoded to obtain $z_i$.
Then, the decoder $\decoder$ is executed top-down to reconstruct the image.
Starting with a constant $r_0$, the decoder iteratively updates $r_i$ using $z_i$ and $e_i$ in each stage, as shown in the right pane of \cref{fig:cloc_method}.
In each stage and the final layer, residual layers and upsampling layers are applied to restore the image to its original resolution.

\textbf{Rate-conditional network layers:}
\cref{fig:cloc_method_layer} shows the details of each layer (\ie, residual block) in our model.
Our model employs the ConvNeXt module~\cite{liu2022convnext} as the basic building blocks.
To achieve variable-rate compression, we adopt the conditional convolution technique~\cite{choi2019var_rate_conditional_ae, duan2023qarv}, which applies an adaptive affine transformation to the convolutional layer output (after layer normalization) to control the rate based on the input $\lambda$.

\textbf{Variable-rate training:}
The (pre-)training objective is to minimize the standard R-D loss for variable rate compression (\cref{eq:cloc_background_rd_loss_var_rate}), except that the rate term consists of the sum of the rates for all latent variables:
\begin{align}
    \label{eq:cloc_method_arch_loss}
    \min & \,  \mathbb{E}_{X, \Lambda} \left[ 
        \sum_{i=1}^{N} - \log_2 p_i(Z_i | \Lambda) + \Lambda \cdot d(X, \hat{X})
    \right],
\end{align}
where $X$ follows the training data distribution, and $\Lambda$ follows $p_\Lambda$, a continuous probability distribution that controls the sampling strategy of $\lambda$ during training.
After the pre-training phase, we freeze $p_Z$ and apply our knowledge replay training strategy presented in \cref{sec:cloc_method_knowledge_replay} to achieve backward-compatible fine-tuning.



\section{Experiments}
\label{sec:cloc_exp}

Our experiments compare various fine-tuning strategies as well as various model architectures for continual learning of image compression.
We begin by describing the setup (\cref{sec:cloc_exp_setup}) and baseline methods (\cref{sec:cloc_exp_baselines}).
Then, \cref{sec:cloc_exp_baseline_comparison} presents the main experimental results for our proposed methods.
Finally, we provide additional experiments to analyze the effectiveness of knowledge replay (\cref{sec:cloc_exp_study_replay}) and our model architecture (\cref{sec:cloc_exp_study_model}).

\subsection{Experiment setup}
\label{sec:cloc_exp_setup}

We consider two continual learning scenarios in image compression: data-incremental learning and rate-incremental learning.
In the former, pre-trained models are fine-tuned on a new dataset, and in the latter, models are fine-tuned with a larger rate range (either going higher or lower).
Detailed configurations are shown in \cref{table:cloc_exp_three_settings}, and the datasets and metrics used are listed below.

\textbf{Datasets:} we use the COCO~\cite{lin2014coco} dataset \textit{train2017} split for pre-training all models.
The dataset contains 118,287 images with around $640\times420$ pixels. We randomly crop the images to $256 \times 256$ patches during training.
For data-incremental learning, we adopt the CelebA-HQ dataset~\cite{lee2020celeba_hq} at $256 \times 256$ pixels, a commonly-used human face image dataset for generative image modeling~\cite{ho2020ddpm}. The dataset consists of 30,000 images, where 24,000 are for training, 3,000 for validation, and the remaining 3,000 for testing.

\textbf{Metrics:} We use bits per pixel (bpp), peak signal-to-noise ratio (PSNR, computed for the RGB space), and BD-Rate~\cite{bjontegaard2001bdrate} to measure compression performance, all of which are standard metrics for image compression.
As described in \cref{sec:cloc_problem}, we evaluate each method for two objectives:
\begin{itemize}
    \item \textbf{Backward compatibility:} We use the fine-tuned model to decode $b_\text{test}^{(0)}$, the bitstreams encoded by the pre-trained model, to obtain reconstructions $\hat{X}_\text{test}^{(0)}$. The bpp is computed for $b_\text{test}^{(0)}$ (which is a constant independent of fine-tuning strategies), and the PSNR is computed between the reconstructions $\hat{X}_\text{test}^{(0)}$ and the original images $X_\text{test}^{(0)}$.
    \item \textbf{New data \& rate performance:} We compress the new data $X_\text{test}^{(1)}$ with the new rates, determined by the $\lambda$ value range $[\lambda_\text{low}^{(1)}, \lambda_\text{high}^{(1)}]$, to obtain bpp and PSNR metrics.
\end{itemize}
We report all results in terms of BD-Rate in the main paper (due to space constraints), and we provide the corresponding PSNR-bpp curves in the Appendix.

\begin{table}
    \setlength{\tabcolsep}{0.5em}
    \footnotesize
    \centering
    \begin{tabular}{l|ccc}
        \hline
            &  $[\lambda_\text{low}^{(0)}, \lambda_\text{high}^{(0)}]$ & $X_\text{train}^{(0)}$ & $X_\text{test}^{(0)}$ \\ \hline
        Pre-training    & $[32, 1024]$  & COCO  & Kodak \\ \hline \hline
            &  $[\lambda_\text{low}^{(1)}, \lambda_\text{high}^{(1)}]$ & $X_\text{train}^{(1)}$ & $X_\text{test}^{(1)}$ \\ \hline
        Data-incremental            & $[32, 1024]$  & CelebA& CelebA\\
        Rate-incremental (low $\to$ high) & $[32, 4096]$  & COCO  & Kodak \\
        Rate-incremental (high $\to$ low) & $[4, 1024]$  & COCO  & Kodak \\
        \hline
    \end{tabular}
    \vspace{-0.2cm}
    \caption{Experiment configurations. We start with a pre-trained model (\textit{pre-training}) and fine-tune it either with new data (\textit{data-incremental}) or new rates (\textit{rate-incremental}).}
    \label{table:cloc_exp_three_settings}
\end{table}

\subsection{Methods in comparison}
\label{sec:cloc_exp_baselines}

In addition to our model, we choose \textit{Mean \& Scale Hyperprior}~\cite{minnen2018joint} (MSH) and \textit{Gaussian Mixture \& Attention}~\cite{cheng2020cvpr} (GMA) as two base models for our experiments.
These two are commonly used and representative models for learned image compression, and we believe the experimental conclusions based on them can be generalized to other existing models.
Since variable-rate compression is required to perform rate-incremental learning, we construct a variable-rate version for each of them and train it in the same way as for our model. The resulting models are referred to as \textit{MSH-VR} and \textit{GMA-VR}, respectively.
\cref{sec:cloc_app_train_details} in the Appendix provides pre-training and fine-tuning hyperparameters, and \cref{sec:cloc_app_vr_baselines} in the Appendix provides details on the variable-rate baselines compared to their fixed-rate models.

We apply the following fine-tuning strategies for each model and compare their performance:
\begin{itemize}
    \item \textbf{Pre-traiend model}: Using the pre-trained model without fine-tuning with new data or rates is the simplest baseline.
    \item \textbf{Fine-tuning the encoder only (FT Enc.)}: We fine-tune the encoder with new data while keeping other parameters frozen. Since the entropy model and the decoder are never changed, backward compatibility is guaranteed.
    \item \textbf{Fine-tuning both the encoder and decoder (FT Enc. \& Dec.)}: We fine-tune all model parameters except for the entropy model parameters. Since the decoder changes, backward compatibility may be lost.
    \item \textbf{Our approach: knowledge replay (KR).} We fine-tune the model's encoder and decoder with the proposed knowledge replay (KR) strategy applied.
\end{itemize}

\begin{table}
    \setlength{\tabcolsep}{0.5em}
    \footnotesize
    \centering
    \begin{tabular}{l|c:c:c} \hline
                            & \multicolumn{3}{c}{BD-Rate (\%) w.r.t. VTM 22.0 $\downarrow$}\\ \hline
                            &  \makecell{Old bitstreams \\ (Kodak)}&  \makecell{New data \\ (CelebA)}&  Avg.\\  \hline
        MSH-VR, pre-trained             & 26.7  & 17.1  & 21.9 \\
        MSH-VR w/ FT Enc.               & 26.7  & 14.8  & 20.8 \\
        MSH-VR w/ FT Enc. \& Dec.       & 419.1 & \textbf{8.95}  & 214.0 \\ \hdashline
        MSH-VR w/ KR (ours)             & \textbf{20.1}  & 9.57  & \textbf{14.8} \\
        \hline
        GMA-VR, pre-trained             & 4.43  & -9.23 &  -2.40 \\
        GMA-VR w/ FT Enc.               & 4.43  & -10.5 &  -3.04\\
        GMA-VR w/ FT Enc. \& Dec.       & 354.6 & \textbf{-17.3} & 168.7\\ \hdashline
        GMA-VR w/ KR (ours)             & \textbf{4.33}  & -14.4 & \textbf{-5.04} \\
        \hline
        Our model, pre-trained          & 1.87  & -13.2 & -5.67\\
        Our model w/ FT Enc.            & 1.87  & -14.6 & -6.37\\
        Our model w/ FT Enc. \& Dec.    & 262.9 & \textbf{-19.0} & 122.0\\ \hdashline
        Our model w/ KR                 & \textbf{0.87}  & -16.6 & \textbf{-7.87}\\ \hline
    \end{tabular}
    \vspace{-0.2cm}
    \caption{Data-incremental learning (COCO $\to$ CelebA) results. PSNR-bpp curves are provided in Appendix, \cref{fig:cloc_appendix_data_inc_rd}.}
    \label{table:cloc_exp_data_inc}
\end{table}

\subsection{Experimental results}
\label{sec:cloc_exp_baseline_comparison}

\textbf{Data-incremental learning}.
\cref{table:cloc_exp_data_inc} show the results (pre-trained on COCO, fine-tuned on CelebA).
We begin by comparing the fine-tuning strategy for the \textit{MSH-VR} model.
Firstly, fine-tuning the encoder (\textit{FT Enc.}) does not provide a significant improvement for new data BD-Rate (17.1\% $\to$ 14.8\%), which is expected since fine-tuning the encoder only reduces the amortization gap~\cite{yang2020improving} and does not improve the model's capacity.
When fine-tuning both the encoder and decoder (\textit{FT Enc. \& Dec.}), the new data BD-Rate is improved by a much larger margin (17.1\% $\to$ 8.95\%), indicating the importance of updating the decoder.
However, this came at the cost of backward compatibility, as shown by the significant increase in BD-Rate for old bitstreams (26.7\% $\to$ 419.1\%).
This indicates that, while the decoder fits the new data well, it becomes incompatible with the old bitstreams.
With our knowledge replay strategy (\textit{MSH-VR w/ KR}), we are able to achieve a competitive new data performance (9.57\% BD-Rate) without sacrificing backward compatibility.
Notably, fine-tuning with our strategy also improves the performance on old bitstreams, which is not the case for the other strategies.
On average, the knowledge replay strategy clearly outperforms the other ones.
These observations are consistent with the results for \textit{GMA-VR} and our model, demonstrating the effectiveness of knowledge replay in data-incremental learning.
Also, when comparing different model architectures, our model achieves the best performance overall in terms of all metrics.

\begin{table}
    \setlength{\tabcolsep}{0.2em}
    \footnotesize
    \centering
    \begin{tabular}{l|c:c:c} \hline
                                & \multicolumn{3}{c}{Kodak BD-Rate (\%) w.r.t. VTM 22.0 $\downarrow$}\\ \hline
                               &  \makecell{Old bitstreams: \\ bpp $\approx$ (0.1,0.9)} &  \makecell{New rate: \\ bpp $\approx$ (0.1,1.6)} &Avg.\\ \hline
        MSH-VR, pre-trained             & 26.7          & -             & -             \\
        MSH-VR w/ FT Enc. \& Dec.       & 282.2         & 23.4          & 152.80        \\ \hdashline
        MSH-VR w/ KR (ours)             & \textbf{19.6} & \textbf{17.3} & \textbf{18.45}\\ \hline
        GMA-VR, pre-trained             & 4.43          & -             & -             \\
        GMA-VR w/ FT Enc. \& Dec.       & 64.1          & 4.56          & 34.33         \\ \hdashline
        GMA-VR w/ KR (ours)             & \textbf{3.39} & \textbf{2.34} & \textbf{2.87} \\ \hline
        Our model, pre-trained          & 1.87          & -             & -             \\
        Our model w/ FT Enc. \& Dec.    & 17.7          & 1.45          & 9.58          \\ \hdashline
        Our model w/ KR                 & \textbf{0.96} & \textbf{0.70} & \textbf{0.83} \\ \hline
    \end{tabular}
    \vspace{-0.2cm}
    \caption{Rate-incremental learning (low $\to$ high) results. PSNR-bpp curves are provided in Appendix, \cref{fig:cloc_appendix_rate_inc_rd}.}
    \label{table:cloc_exp_rate_inc}
\end{table}
\begin{table}
    \setlength{\tabcolsep}{0.2em}
    \footnotesize
    \centering
    \begin{tabular}{l|c:c:c} \hline
                                & \multicolumn{3}{c}{Kodak BD-Rate (\%) w.r.t. VTM 22.0 $\downarrow$}\\ \hline
                               &  \makecell{Old bitstreams: \\ bpp $\approx$ (0.1,0.9)} &  \makecell{New rate: \\ bpp $\approx$ (0.03,0.9)} &Avg.\\ \hline
        MSH-VR, pre-trained             & 26.7          & -             & -             \\
        MSH-VR w/ FT Enc. \& Dec.       & 35.1          & 37.6          & 36.4          \\ \hdashline
        MSH-VR w/ KR (ours)             & \textbf{18.9} & \textbf{29.9} & \textbf{24.4} \\ \hline
        GMA-VR, pre-trained             & 4.43          & -             & -             \\
        GMA-VR w/ FT Enc. \& Dec.       & 10.0          & 9.11          & 9.56          \\ \hdashline
        GMA-VR w/ KR (ours)             & \textbf{1.75} & \textbf{6.92} & \textbf{4.34} \\ \hline
        Our model, pre-trained          & 1.87          & -             & -             \\
        Our model w/ FT Enc. \& Dec.    & 14.33         & 5.18          & 9.76          \\ \hdashline
        Our model w/ KR                 & \textbf{0.86} & \textbf{4.26} & \textbf{2.56} \\ \hline
    \end{tabular}
    \vspace{-0.2cm}
    \caption{Rate-incremental learning (high $\to$ low) results. PSNR-bpp curves are provided in Appendix, \cref{fig:cloc_appendix_rate_inc_low_rd}.}
    \label{table:cloc_exp_rate_inc_low}
\end{table}

\textbf{Rate-incremental learning}.
\cref{table:cloc_exp_rate_inc} presents the results for rate-incremental learning (from low rates to higher rates).
In rate-incremental experiments, we omit the \textit{FT Enc.} baseline, since fine-tuning the encoder alone cannot effectively extend the rate range of any considered models (see Appendix, \cref{sec:cloc_app_rate_inc_exp} for details).
Starting with the \textit{MSH-VR} model, we observe that fine-tuning both the encoder and decoder (\textit{FT Enc. \& Dec.}) is able to extend the operational rate range of the model with a similar BD-Rate w.r.t. VTM for the new rates.
However, the performance on old bitstreams is significantly degraded, similar to the observation in data-incremental learning experiments.
By applying our knowledge replay strategy (\textit{MSH-VR w/ KR}), in contrast, the model is able to achieve a competitive BD-Rate for the new rates (17.3\%) while maintaining backward compatibility.
Again, the results for \textit{GMA-VR} and our model show a similar pattern.
Overall, our model with KR outperforms the baselines, validating its effectiveness in rate-incremental learning.
\zhihao{
For rate-incremental learning from high rates to low rates (\cref{table:cloc_exp_rate_inc_low}), the above observations stay the same.
}


\subsection{Experimental analysis: knowledge replay}
\label{sec:cloc_exp_study_replay}

The effectiveness of the proposed knowledge replay strategy has already been verified in the previous experiments.
We now provide additional experiments to answer the following questions that aim to analyze the individual components of our knowledge replay strategy.

\begin{table}
    \setlength{\tabcolsep}{0.5em}
    \centering
    \footnotesize
    \begin{tabular}{c|cc|ccc}
        \hline
                &  \multicolumn{2}{c|}{Fine-tuning} &  \multicolumn{3}{c}{BD-Rate w.r.t. VTM 22.0}    \\
         Config.& Data      &  KR loss  &  Old bits.    &  New data     & Avg.  \\ \hline
         0      & -         & -         &  1.87         &  -13.2        & -5.67 \\ \hdashline
         1      & CA        &           &  262.9        & \textbf{-19.0}& 121.9 \\
         2      & CA + COCO &           &  23           & -16.9         & 3.05  \\ \hdashline
         3      & CA        &\checkmark &  3.67         &  -16.3        & -6.32 \\
         4 (ours)& CA + COCO&\checkmark &  \textbf{0.87}&  -16.6        & \textbf{-7.87}\\ \hline
    \end{tabular}
    \vspace{-0.2cm}
    \caption{
        Ablative analysis of our knowledge replay-based training strategy for data-incremental learning (COCO $\to$ CelebA).
        For the \textit{``data''} column, \textit{CA} denotes the CelebA dataset.
    }
    \label{table:cloc_exp_kr_ablation}
\end{table}

\zhihao{
\textbf{What contributes to the backward compatibility?}
In data-incremental learning, there are two components in our knowledge replay strategy that may help backward compatibility: the replayed training data and the knowledge replay loss.
To analyze the contribution of each component, we freeze the entropy model parameters of our model and fine-tune it with the two components separately.
To analyze the contribution of each component, we start from ``\textit{Our model w/ FT Enc. \& Dec.}'' and apply the two components one by one.
\cref{table:cloc_exp_kr_ablation} shows the results.
\textit{Config. 0} is the pre-trained model, and \textit{Config. 1} is the \textit{``FT Enc. \& Dec.''} baseline that does not retain backward compatibility.
With replayed data (\textit{Config. 2}), backward compatibility is largely improved (262.9\% $\to$ 23\% BD-Rate) but is still worse than the pre-trained model.
When the knowledge replay loss is applied (\textit{Config. 3} and \textit{Config. 4}), the performance on old bitstreams becomes comparable to the pre-trained model.
We conclude that both the replayed data and the loss function contribute to backward compatibility, and the knowledge replay loss is more important among the two.
}

\begin{table}[t]
    \footnotesize
    \centering
    \begin{tabular}{l|ccccc}
    \hline
         $\alpha$                   &  0.0  &  0.25 &  0.5  &  0.75 & 1.0   \\ \hline
         BD-rate: old bitstreams    & 262.9 & 1.40  & 0.87  & 0.72  & \textbf{0.63}  \\
         BD-rate: new data          & \textbf{-19.0} & -16.8 & -16.6 & -16.3 & -14.4 \\ \hdashline
         Avg. BD-rate               &  122.0& -7.7  & \textbf{-7.9} &  -7.8 & -6.9  \\
    \hline
    \end{tabular}
    \vspace{-0.2cm}
    \caption{Data-incremental learning (COCO $\to$ CelebA) results for our model with varying $\alpha$. The scalar $\alpha \in [0, 1]$ controls the ratio of replayed data in fine-tuning, where $\alpha = 0.0$ means no replay, and $\alpha = 1.0$ means no new data. BD-rate is w.r.t. VTM 22.0.}
    \label{table:cloc_exp_alpha}
\end{table}

\textbf{What is the impact of $\alpha$ in the knowledge replay loss function (\cref{eq:cloc_method_kr_loss})?}
We train our model for data-incremental learning with varying $\alpha$, the scalar that controls the ratio of replayed data in each training iteration.
Results are shown in \cref{table:cloc_exp_alpha}.
Firstly, it is clear that as $\alpha$ grows from 0 to 1, the performance on old bitstreams monotonically improves (\ie, the BD-Rate decreases).
Recall that when $\alpha = 0$, no replay is performed, and the model is trained with only the new data; when $\alpha = 1$, the model is trained with only the replayed data.
The results are thus consistent with the intuition that more replay leads to better backward compatibility.
When it comes to the new data performance, the trend is reversed: as $\alpha$ grows, the performance on new data monotonically degrades, which again matches the intuition.
On average, the performance is comparable for $\alpha \in [0.25, 0.75]$.
We conclude that our approach is insensitive to the choice of $\alpha$ and can achieve a good trade-off between backward compatibility and new data performance.
Our experiments choose $\alpha = 0.5$ by default, but in practice, the choice of $\alpha$ can be treated as a hyperparameter and determined by the application requirements.

\subsection{Experimental analysis: model architecture}
\label{sec:cloc_exp_study_model}

\begin{table}[t]
    \setlength{\tabcolsep}{0.44em}
    \scriptsize
    \centering
    \begin{tabular}{l|c|cc|cc|cc}
    \hline
        & & \multicolumn{6}{c}{Latency (in seconds)} \\ \hline
                    & Params. &  \multicolumn{2}{c|}{Entropy Coding}&\multicolumn{2}{c|}{Network (CPU)}&\multicolumn{2}{c}{Network (GPU)}\\
                    &       &  Enc. & Dec.  & Enc.  &  Dec. &  Enc. & Dec.  \\ \hline
        MSH + VR    & 19.2M & 0.026 & 0.068 & 0.318 & 0.325 & 0.004 & 0.010 \\
        GMA + VR    & 33.4M & 3.072 & 6.302 & 0.941 & 1.221 & 0.006 & 0.022 \\ \hdashline
        Ours        & 35.5M & 0.046 & 0.052 & 0.474 & 0.393 & 0.026 & 0.011 \\ \hline
    \end{tabular}
    \begin{tablenotes}
        \scriptsize
        \item *Hardware: Intel 10700K CPU (using four threads) and Nvidia Quadro 6000 GPU.
        \item *Latency is the average time to encode/decode a Kodak image (768$\times$512 pixels), averaged over all 24 images. Time includes entropy coding.
    \end{tablenotes}
    \vspace{-0.2cm}
    \caption{Computational complexity of the model architectures used in our experiments.}
    \label{tab:cloc_exp_complexity}
\end{table}

\textbf{Computational complexity.}
\cref{tab:cloc_exp_complexity} shows the computational complexity of our model and the two baselines.
The metrics include the number of parameters, entropy coding latency, and neural network forward pass latency.
Except for a few exceptions (the parameter count of \textit{MSH-VR}, the entropy coding latency of \textit{GMA-VR}, and the GPU encoding latency of our model), all methods are mostly comparable in terms of computational complexity.
Thus, we conclude that the performance improvement of our model is not due to the increase in computational complexity.

\begin{table}
    \setlength{\tabcolsep}{0.5em}
    \footnotesize
    \centering
    \begin{tabular}{l|c:c:c} \hline
                            & \multicolumn{3}{c}{BD-Rate (\%) w.r.t. VTM 22.0 $\downarrow$}\\ \hline
                            &  \makecell{Old bitstreams \\ (Kodak)}&  \makecell{New data \\ (CelebA)}&  Avg.\\  \hline
        Sequential $p_Z$ and $\decoder$ & 4.42  & -12.8 & -4.19 \\
        Sequential $p_Z$ and $\decoder$ w/ KR          & 3.18  & -15.8 & -6.31 \\ \hdashline
        Parallel $p_Z$ and $\decoder$   & 1.87  & -13.2 & -5.67 \\
        Parallel $p_Z$ and $\decoder$ w/ KR & \textbf{0.87}  & \textbf{-16.6} & \textbf{-7.87}\\ \hline
    \end{tabular}
    \vspace{-0.2cm}
    \caption{Comparing architecture variants of our model in terms of data-incremental learning (COCO $\to$ CelebA).}
    \label{table:cloc_exp_2b_vs_1b}
\end{table}

\textbf{Impact of decoupling the entropy model and the decoder.}
Recall that in order to reduce the parameters of $p_Z$, our architecture decouples $p_Z$ and $\decoder$ into two parallel branches.
To analyze the impact of doing so, we construct a variant of our model where $p_Z$ and $\decoder$ are executed in sequential (like in Hyperprior-based models; see \cref{fig:cloc_problem_hyperprior}).
For a fair comparison, the sequential version has the same number of latent feature channels and the same number of parameters as the parallel model, and training is performed in the same way as for all previous experiments.
We show the data-incremental learning results in \cref{table:cloc_exp_2b_vs_1b}.
We observe that our model (\textit{Parallel $p_Z$ and $\decoder$ w/ KR}) achieves a better performance than the sequential version on both old bitstreams and new data, which verifies our design.


\subsection{Discussion}

Our experiments show that neural image compressors can adapt to new data and rates in a backward-compatible manner by using the proposed training strategy.
In addition to continual learning applications, this observation offers insights into related research such as the standardization of learned image compression.
Despite recent attempts toward this goal~\cite{ma2022tpami_iwave, ascenso2023jpeg_ai_standard}, it remains an open question about which part of a selected neural compressor needs to be standardized.
Our findings suggest a possible direction: only the entropy model (instead of the entire model architecture and parameters) needs to be standardized, and other components (\eg, the decoder network) could be fine-tuned over time with backward-compatible training strategies.


\section{Conclusion}
This paper presents two approaches: a knowledge replay-based training strategy and a neural network architecture, for continual learning of image compression.
Our knowledge replay strategy enables existing compressors to adapt to new data and target rates while ensuring that previously compressed bitstreams remain decodable.
Through extensive experiments, we conclusively answer the question raised at the beginning of the paper: neural network-based image compressors can be learned continually in a backward-compatible manner, achieving improved performance on new data, new rates, and old bitstreams.

\textbf{Limitations and future work.} Our work serves as a preliminary study on continual learning for image compression.
Despite its effectiveness, our knowledge replay strategy assumes unconstrained training resources, which may not be true in practice.
Also, we focus on the decoder's backward compatibility, while the same problem can be studied for the encoder.
For future work, a possible direction is to extend our two-step method (pre-training and fine-tuning) to multi-step continual learning scenarios.

{
    \footnotesize
    \bibliographystyle{ieeenat_fullname}
    \bibliography{references}
}

\clearpage
\setcounter{page}{1}
\maketitlesupplementary

\section{Appendix: Model Architecture Details}
\label{sec:cloc_app_model_arch}
In the main paper, \cref{sec:cloc_method_architecture} provides a high-level overview of the proposed model architecture.
This section provides more details about the model architecture, such as the number of channels and stride sizes for each layer.

\begin{figure*}[t]
    \centering
    \includegraphics[width=\linewidth]{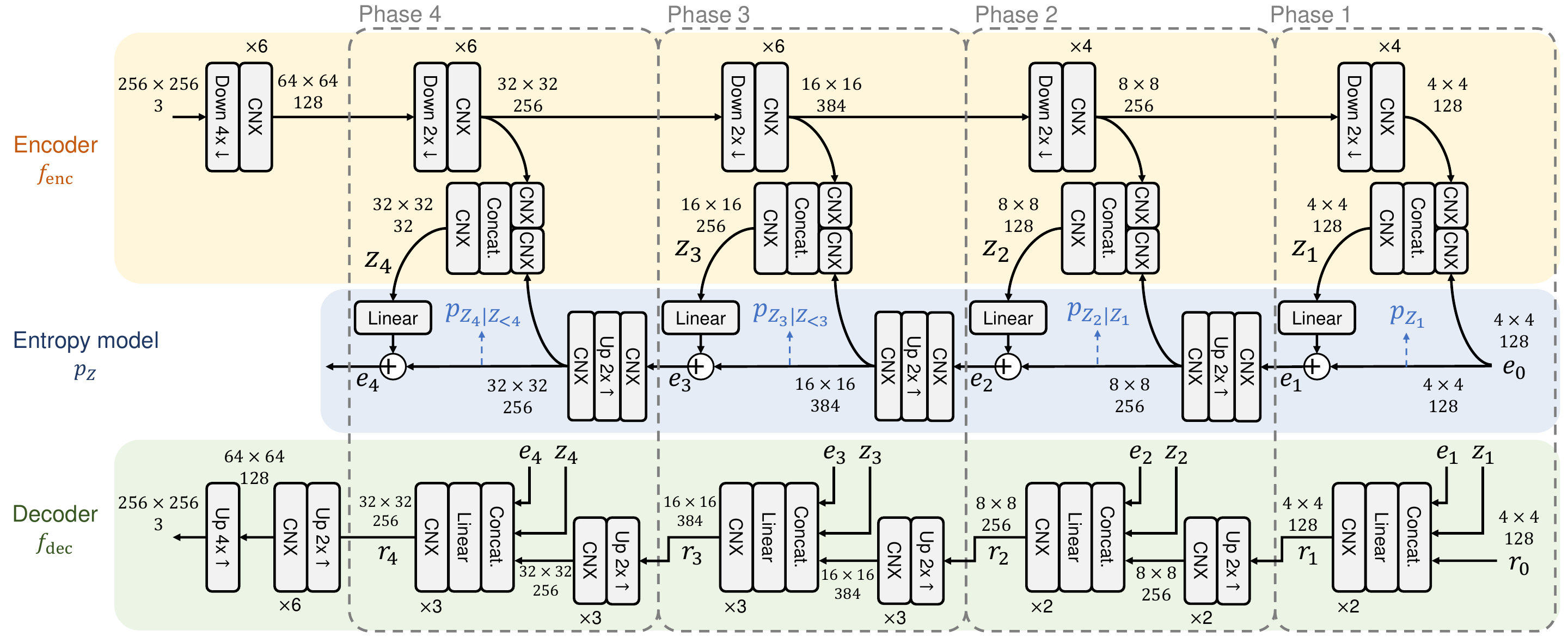}
    \caption{
        Detailed architecture of the proposed model. In the figure, \textit{CNX} denotes a ConvNeXt block~\cite{liu2022convnext} conditioned on lagrange multiplier $\lambda$, as described in \cref{fig:cloc_method_layer}.
        Dimensionality of the layer outputs are shown in the format of \textit{height $\times$ width} and \textit{channels}, where the spatial dimensions (height and width) are for a $256 \times 256$ input image, and they scales linearly with the input image size.
    }
    \label{fig:cloc_appendix_arch_details}
\end{figure*}

The detailed model architecture is shown in \cref{fig:cloc_appendix_arch_details}, where the model components are marked in the same way as in \cref{sec:cloc_method_architecture}.
Our model contains four phases, all of which have the same structure, while only different in (1) the number of feature channels, (2) the number of ConvNeXt blocks, and (3) the first phase starts from bias $e_0$ and $r_0$ instead of the feature maps from the previous phase.

The spatial dimensions (height and width) in the figure are for an input image with $256 \times 256$ pixels.
Since the model is fully convolutional, the spatial dimensions of intermediate layer outputs scales accordingly with the input image size.
Both initial bias features $e_0$ and $r_0$ have a shape of $1 \times 1 \times 128$, and they are repeated spatially to match the spatial dimensions of $z_1$.

\section{Appendix: Training and Fine-tuning Details}
\label{sec:cloc_app_train_details}

\cref{table:qarv_appendix_hyp_param} lists the pre-training and fine-tuning hyperparameters used in our experiments.
For a fair comparison, we use the same hyperparameters for training all models, including our proposed model and the baseline models (\,  i.e., \textit{MSH-VR} and \textit{GMA-VR}).
Note that the fine-tuning dataset varies for different sets of experiments. For data-incremental learning, we use CelebA-HQ~\cite{lee2020celeba_hq}, and for rate-incremental learning, we use COCO~\cite{lin2014coco}, which is the same as the pre-training dataset.

\begin{table}[ht]
\small
\centering
\begin{tabular}{l|c|c}
\hline
               & Pre-training       & Fine-tuning         \\ \hline
Data augmentation & Crop, h-flip    & Crop, h-flip        \\
Input size     & 256x256            & 256x256             \\ \hdashline
Optimizer      & Adam               & Adam                \\
Learning rate  & $2 \times 10^{-4}$ & $1 \times 10^{-4}$  \\
LR schedule    & Constant + cosine  & Cosine              \\
Weight decay   & 0.0                & 0.0                 \\ \hdashline
Batch size     & 32                 & 32                  \\
\# iterations  & 500K               & 100K                \\
\# images seen & 16M                & 3.2M                \\ \hdashline
Gradient clip  & 2.0                & 2.0                 \\
EMA            & 0.9999             & -                   \\ \hdashline
GPU            & 1 $\times$ RTX 3090 & 1 $\times$ A40     \\
Time           & $\approx$ 51 hours & $\approx$ 11 hours  \\ \hline
\end{tabular}
\caption{Training Hyperparameters. The GPU time is for training our proposed model, and all other hyperparameters are the same for all models.}
\label{table:qarv_appendix_hyp_param}
\end{table}

\section{Appendix: Variable-Rate Baseline Models}
\label{sec:cloc_app_vr_baselines}

\begin{figure}[t]
    \centering
    \includegraphics[width=0.92\linewidth]{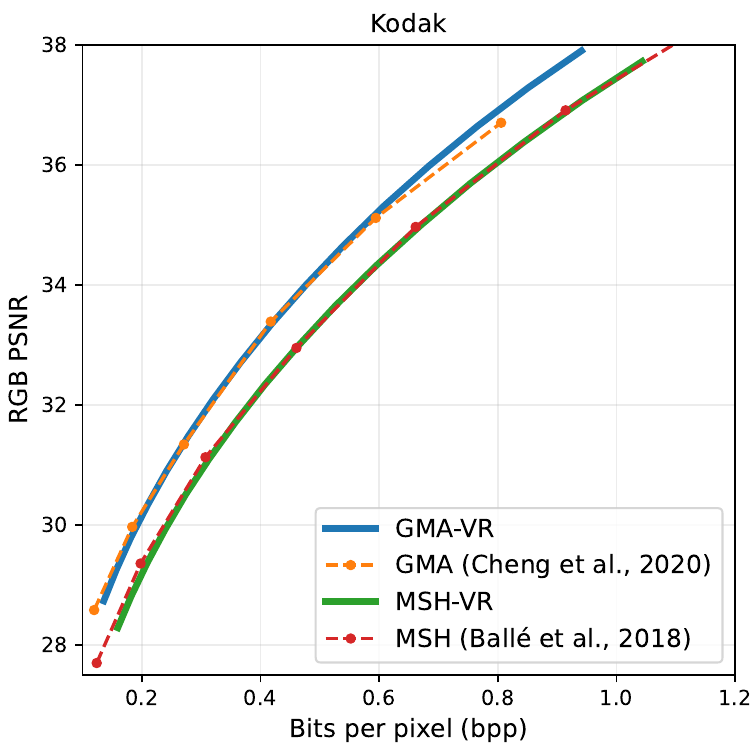}
    \vspace{-0.2cm}
    \caption{The variable-rate version of the baseline models that we constructed (MSH-VR and GMA-VR) are comparable to the original ones (MSH~\cite{minnen2018joint} and GMA~\cite{cheng2020cvpr}) in terms of PNSR-bpp performance on Kodak.}
    \label{fig:cloc_appendix_baseline_vr}
\end{figure}

In the main paper (\cref{sec:cloc_exp_baselines}), we mentioned that we construct variable-rate versions of the two baseline models (\ie, \textit{MSH-VR} and \textit{GMA-VR}) in order to use them in the rate-incremental learning experiment.
\cref{fig:cloc_appendix_baseline_vr} shows the rate-distortion performance of the variable-rate versions compared to the original ones.
As shown in the figure, the variable-rate versions achieve similar performance as the original ones, which validates the our experimental setting.

\section{Appendix: Experimental Results}

\subsection{PSNR-Bpp curves for the main experiments}
Due to the space constraint, we show only BD-rate results without PSNR-bpp curves in the main paper.
This section provides the PSNR-bpp curves for the main experiments (\cref{sec:cloc_exp_baseline_comparison}).

\begin{figure*}[t]
    \centering
    \begin{subfigure}{0.92\linewidth}
        \includegraphics[width=\linewidth]{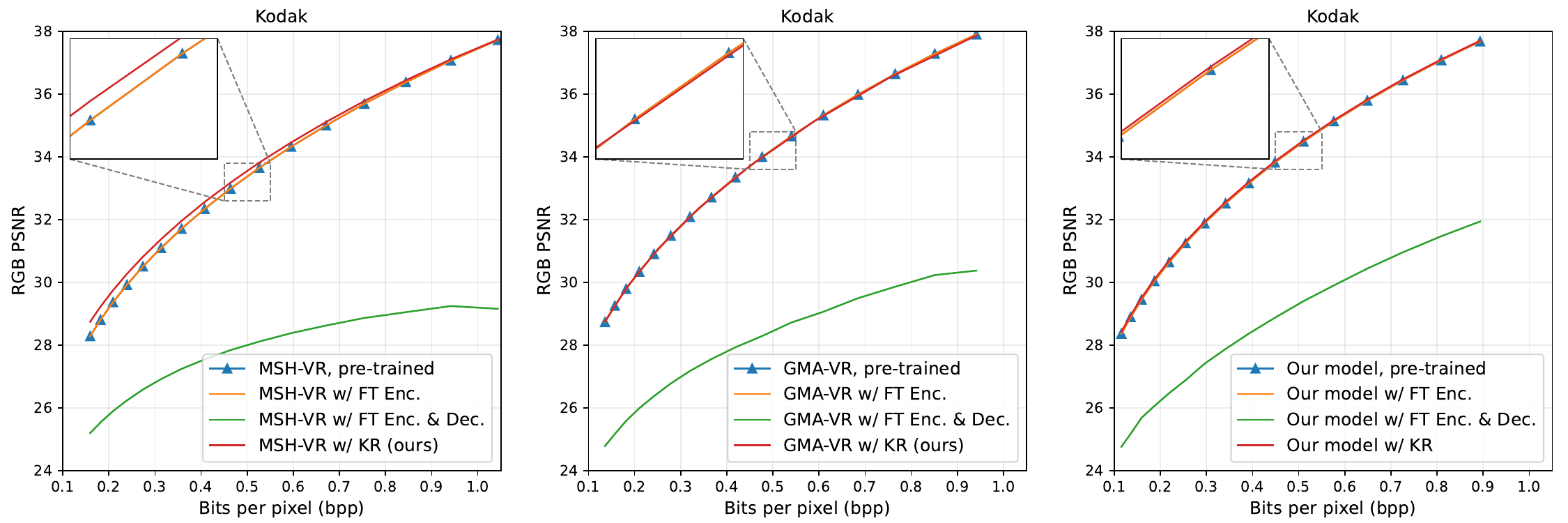}
        \caption{Performance on the old bitsreams of Kodak (backward compatibility). Each subfigure shows the performance of a different model.}
        \label{fig:cloc_appendix_data_inc_rd_old}
    \end{subfigure}
    \begin{subfigure}{0.92\linewidth}
        \includegraphics[width=\linewidth]{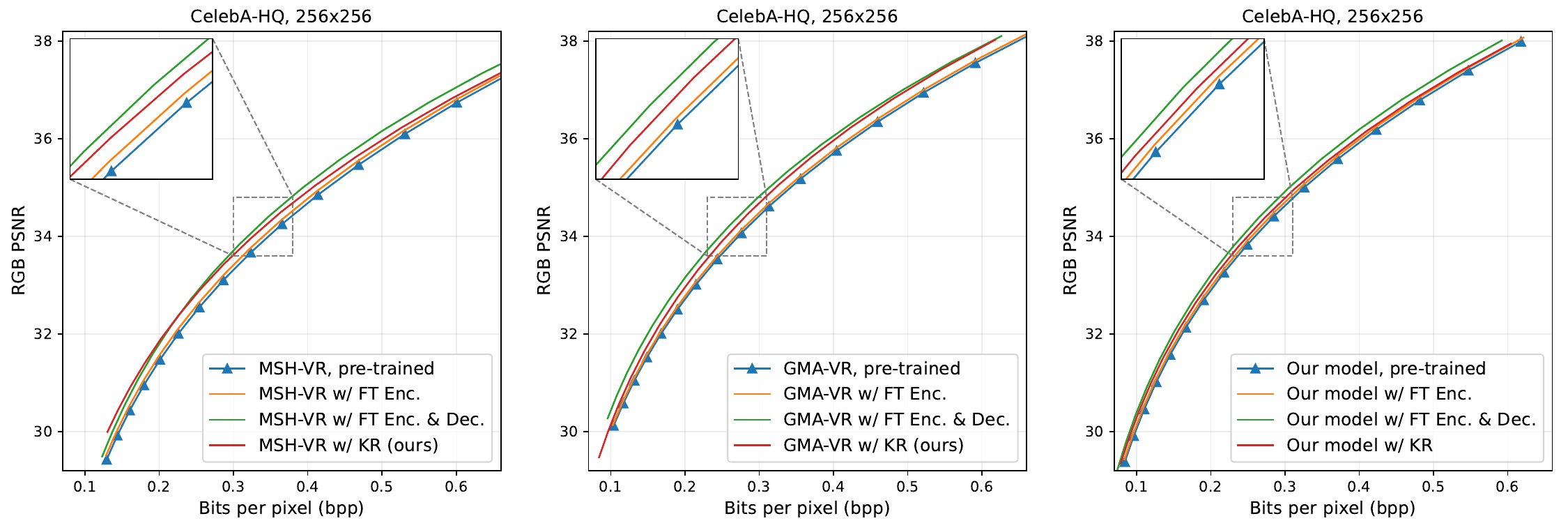}
        \caption{Performance on CelebA-HQ (new-data performance). Each subfigure shows the performance of a different model.}
        \label{fig:cloc_appendix_data_inc_rd_new}
    \end{subfigure}
    \vspace{-0.24cm}
    \caption{PSNR-Bpp curves for \textbf{data-incremental learning} experiments. In figure (a), the \textit{``models, pre-trained''} curves overlap with the \textit{``models w/ FT Enc.''} curves because their decoder are the same.}
    \label{fig:cloc_appendix_data_inc_rd}
\end{figure*}

\cref{fig:cloc_appendix_data_inc_rd} shows the PSNR-bpp curves for data-incremental learning experiments, which includes the backward compatibility experiment (\cref{fig:cloc_appendix_data_inc_rd_old}) and the new-data performance experiment (\cref{fig:cloc_appendix_data_inc_rd_new}).
For backward compatibility, it is clear that models with fine-tuned encoder and decoder suffer a significant performance drop on the old bitstreams, while other fine-tuned models obtain comparable performance as the pre-trained models.
Among them, our proposed knowledge replay strategy achieves even better performance than using the pre-trained model directly.
For new-data performance, our method achieves comparable performance as the models with fine-tuned encoder and decoder (which are not backward compatible), and outperforms the pre-trained models by a clear margin.
These observations are consistent with what we have observed in the BD-rate results in the main paper.

\begin{figure*}[t]
    \centering
    \begin{subfigure}{0.8\linewidth}
        \includegraphics[width=\linewidth]{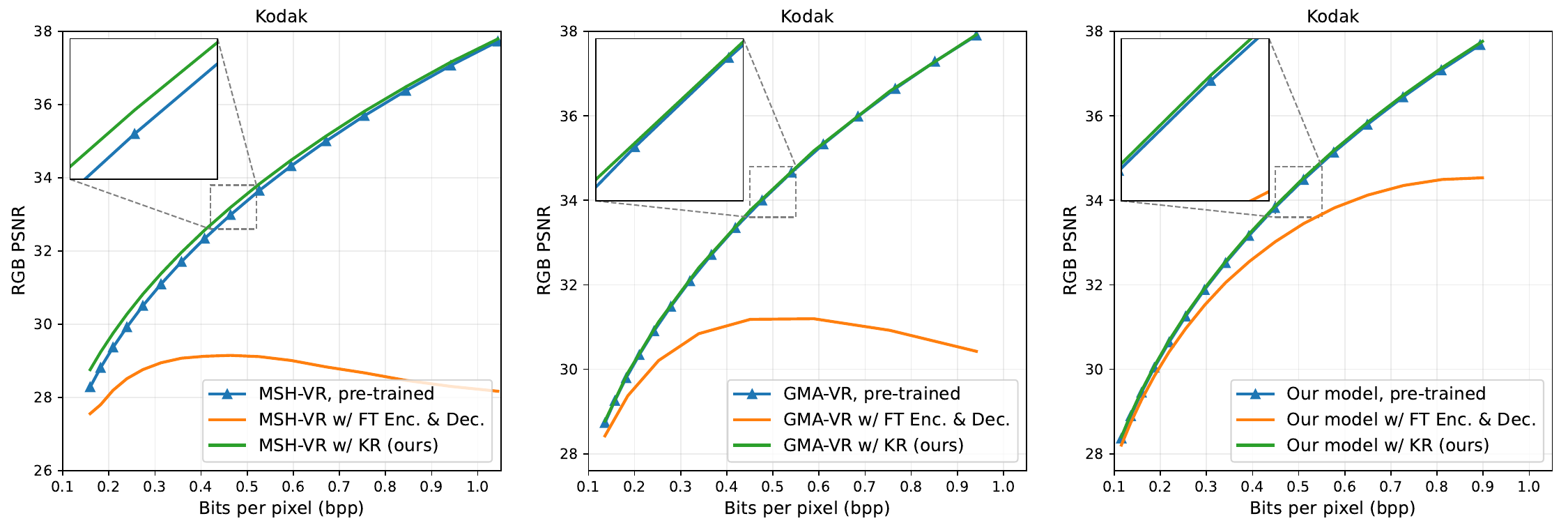}
        \vspace{-0.5cm}
    \caption{Backward compatibility (bpp range is around $[0.1, 0.9]$). Each subfigure shows the performance of a different model.}
    \end{subfigure}
    \begin{subfigure}{0.8\linewidth}
        \includegraphics[width=\linewidth]{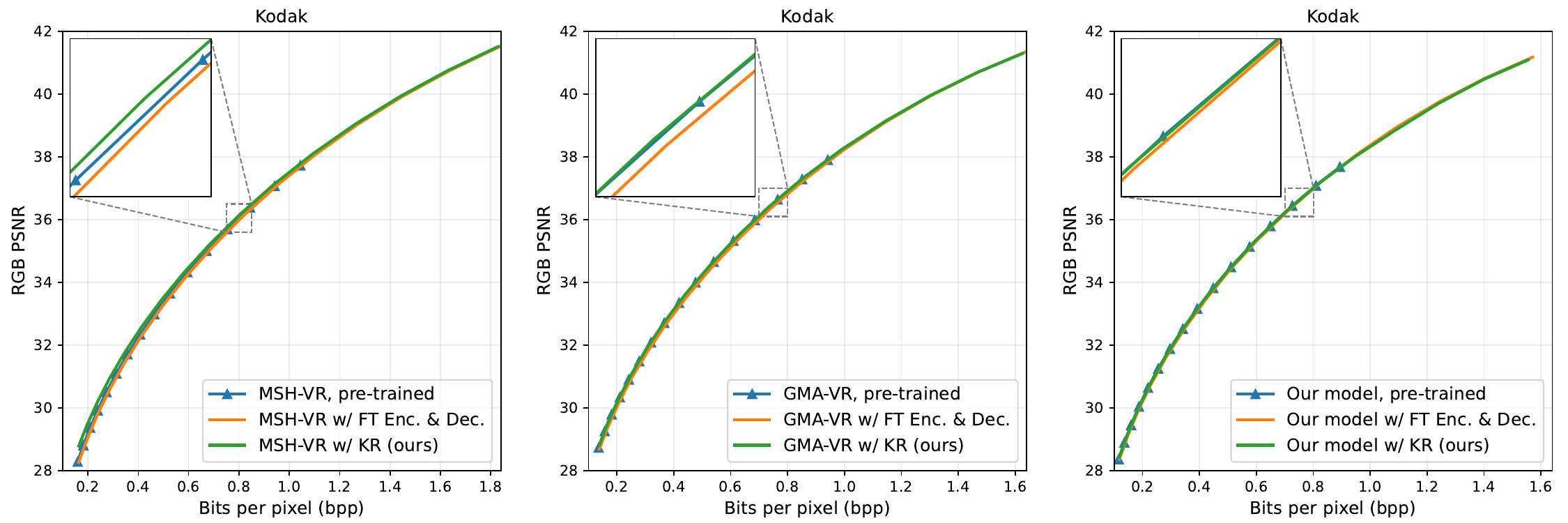}
        \vspace{-0.5cm}
        \caption{New-rate performance (bpp range is around $[0.1, 1.6]$). Each subfigure shows the performance of a different model.}
    \end{subfigure}
    \vspace{-0.32cm}
    \caption{PSNR-Bpp curves for \textbf{rate-incremental learning (low $\to$ high)} experiments.}
    \label{fig:cloc_appendix_rate_inc_rd}
\end{figure*}

\begin{figure*}[t]
    \centering
    \begin{subfigure}{0.8\linewidth}
        \includegraphics[width=\linewidth]{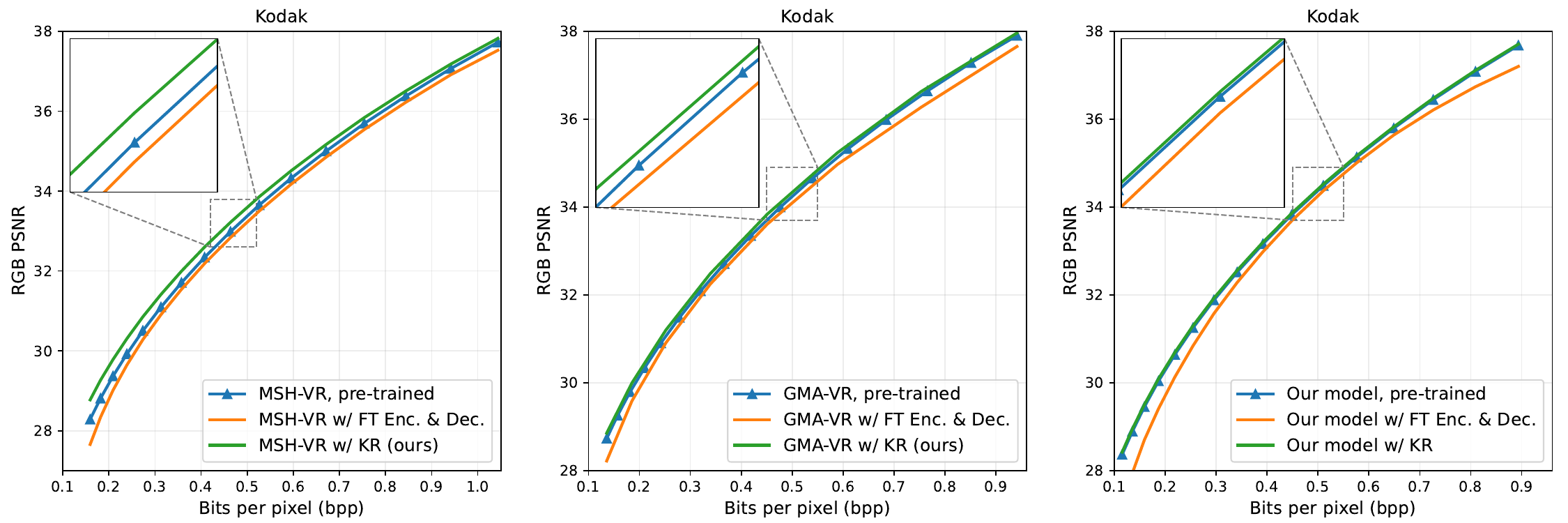}
        \vspace{-0.5cm}
        \caption{Backward compatibility (bpp range is around $[0.1, 0.9]$). Each subfigure shows the performance of a different model.}
    \end{subfigure}
    \begin{subfigure}{0.8\linewidth}
        \includegraphics[width=\linewidth]{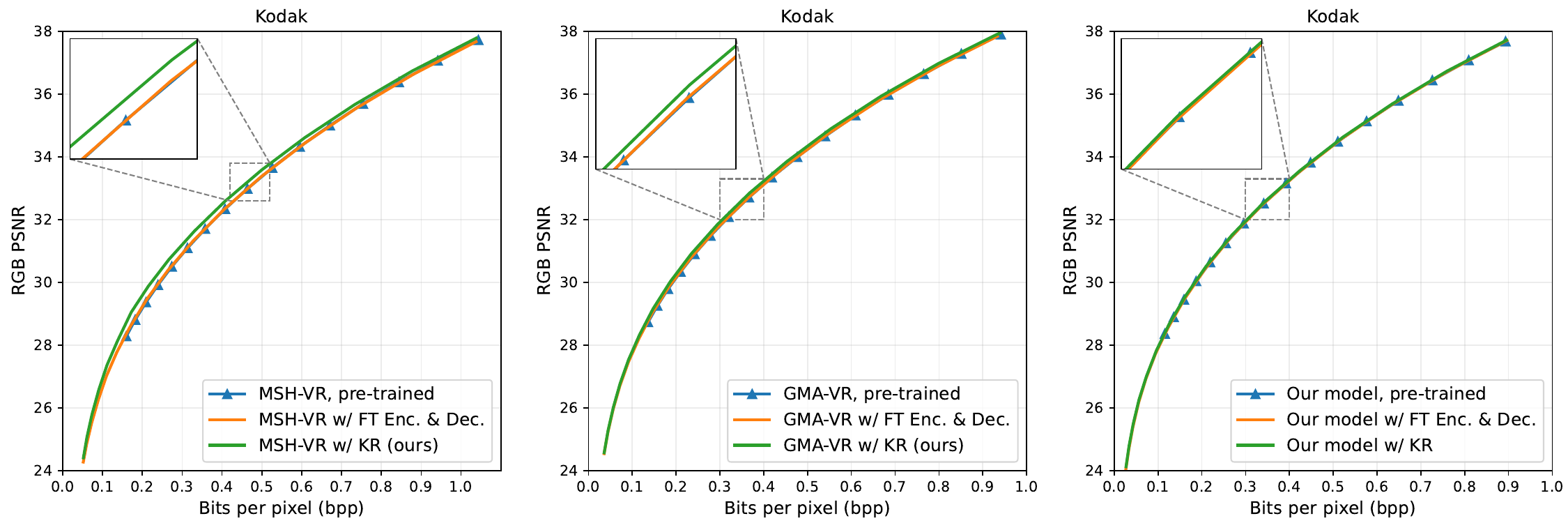}
        \vspace{-0.5cm}
        \caption{New-rate performance (bpp range is around $[0.03, 0.9]$). Each subfigure shows the performance of a different model.}
    \end{subfigure}
    \vspace{-0.32cm}
    \caption{PSNR-Bpp curves for \textbf{rate-incremental learning (high $\to$ low)} experiments.}
    \label{fig:cloc_appendix_rate_inc_low_rd}
\end{figure*}

We show the PSNR-bpp curves for rate-incremental learning experiments, including the \textit{low-to-high} experiment (\cref{fig:cloc_appendix_rate_inc_rd}) and the \textit{high-to-low} experiment (\cref{fig:cloc_appendix_rate_inc_low_rd}).
The results are consistent with previous observations:
(1) Fine-tuning the encoder and decoder does not preserve backward compatibility, while our approach does; and
(2) Our approach even outperforms all other methods in terms of new-rate performance.

\subsection{Fine-tuning the encoder does not generalize the model to new rates}
\label{sec:cloc_app_rate_inc_exp}
\begin{figure}[t]
    \centering
    \includegraphics[width=0.92\linewidth]{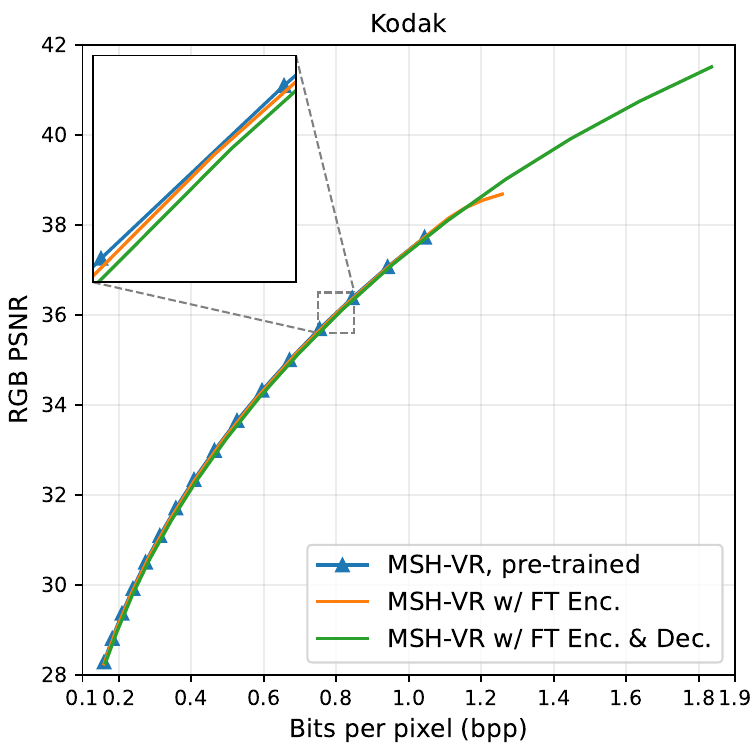}
    \vspace{-0.2cm}
    \caption{Fine-tuning the encoder does not effectively generalize the pre-train model (MSG-VR, for example) to new rates.}
    \label{fig:cloc_appendix_ft_enc_rate}
\end{figure}

We mentioned in \cref{sec:cloc_exp_baseline_comparison} that fine-tuning the encoder alone cannot effectively extend the rate range of the pre-trained models.
We provide an example for showing this in \cref{fig:cloc_appendix_ft_enc_rate}, where we show the rate-incremental learning (low $\to$ high) performance of the pre-trained MSH-VR, the one with fine-tuned encoder (\textit{MSH-VR w/ FT Enc.}), and the one with fine-tuned encoder and decoder (\textit{MSH-VR w/ FT Enc. \& Dec.}).
As shown in the figure, fine-tuning the encoder marginally extends the rate range of the pre-trained model, and the PSNR drops visibly when the rate is higher than maximum rate of the pre-trained model.
Thus, we do not use this strategy in our rate-incremental learning experiments.

\end{document}